\documentclass[]{aa} 

\usepackage{graphicx}
\usepackage{txfonts}
\usepackage{natbib}
\usepackage{gensymb}
\usepackage{amsmath}
\usepackage{dblfloatfix}

\begin{document}

   \title{Hybrid star HD 81817 accompanied by brown dwarf or substellar companion}

   \author{Tae-Yang Bang\inst{1,4},
          Byeong-Cheol Lee\inst{2,3},
          V. Perdelwitz\inst{5},
          Gwanghui Jeong\inst{2,3},
          Inwoo Han\inst{2,3},
          Hyeong-Il Oh\inst{1,2},
          \and
          Myeong-Gu Park\inst{1,4}\fnmsep\thanks{Corresponding author}
          }

   \institute{Department of Astronomy and Atmospheric Sciences, Kyungpook National University,
               80 Daehakro, Bukgu, 41566 Daegu, Korea\\
              \email{[qkdxodid1230;mgp]@knu.ac.kr}
         \and
             Korea Astronomy and Space Science Institute, 776 Daedukdae-ro, Yuseong-gu, 34055 Daejeon, Korea
                 \and
             Korea University of Science and Technology, 217 Gajeong-ro, Yuseong-gu, 34113 Daejeon, Korea
         \and
             Research and Training Team for Future Creative Astrophysicists and Cosmologists (BK21 Plus Program)
         \and
                        Hamburger Sternwarte, Gojenbergsweg 112, 21029, Hamburg, Germany\\
                                           }
   \date{Received 02 September 2019 / Accepted 11 May 2020}

  \abstract
   {HD 81817 is known as a hybrid star. Hybrid stars have both cool stellar wind properties and Ultraviolet (UV) or even X-ray emission features of highly ionized atoms in their spectra. A white dwarf companion has been suggested as the source of UV or X-ray features. HD~81817 has been observed since 2004 as a part of a radial velocity (RV) survey program to search for exoplanets around K giant stars using the Bohyunsan Observatory Echelle Spectrograph  at the 1.8 m telescope of Bohyunsan Optical Astronomy Observatory in Korea. We obtained 85 RV measurements between 2004 and 2019 for HD 81817 and found two periodic RV variations. The amplitudes of RV variations are around 200 m s$^{-1}$, which are significantly lower than that expected from a closely orbiting white dwarf companion. Photometric data and relevant spectral lines were also analyzed to help determine the origin of the periodic RV variations. We conclude that 627.4-day RV variations are caused by intrinsic stellar activities such as long-term pulsations or rotational modulations of surface activities based on $\rm{H_{\alpha}}$ equivalent width (EW) variations of a similar period. On the other hand, 1047.1-day periodic RV variations are likely to be caused by a brown dwarf or substellar companion, which is corroborated by a recent GAIA proper motion anomaly for HD 81817. The Keplerian fit yields a minimum mass of 27.1 $\it M_\mathrm{Jup}$, a semimajor axis of 3.3 AU, and an eccentricity of 0.17 for the stellar mass of 4.3 $M_{\odot}$ for HD 81817. The inferred mass puts HD 81817 b in the brown dwarf desert.}

   \keywords{Binaries (including multiple): close -- stars: binaries: spectroscopic -- stars: activity -- stars: oscillations (including pulsations) -- stars: brown dwarf -- stars: individual: HD 81817 -- Techniques: radial velocities}

\titlerunning{Hybrid star HD 81817 accompanied by brown dwarf or substellar companion}
\authorrunning{Bang et al. (2020)}

   \maketitle

\section{Introduction}

Hybrid stars exhibit both emission features in the high-energy regime \citep{ayres1} as well as cool stellar wind feature emission lines near 1500 {\AA} \citep{Hartmann1980}. These objects were first discovered by \cite{Hartmann1980} and more hybrid stars have been found subsequently (\citealt{Hartmann1981}; \citealt{Reimers82}; \citealt{Reimers84}; \citealt{Haisch}; \citealt{Reimers1992}). The origin of the X-ray emission of some hybrid stars such as $\alpha$ Per \citep{ayres2} and $\gamma$ Dra \citep{ayres3} is suspected to be unresolved foreground or background sources.
 
HD 81817 was suggested to be a hybrid star by \cite{Reimers84} based on the flux distribution in International Ultraviolet Explorer (IUE) Ultraviolet (UV) spectra. The UV spectra show strong emission lines, almost these emission lines are similar to those of another hybrid star, $\theta$ Her (HD 163770). Owing to an apparent UV continuum feature, \cite{Reimers84} furthermore suggested that HD81817 is accompanied by a white dwarf with an effective temperature of $\sim$ 20,000 K. 

Although HD 81817 is also an X-ray source, \cite{bilikova} suspected that the origin of the X-ray emission is not likely to be a white dwarf companion because the X-ray to bolometric luminosity ratio of HD 81817 falls within the range expected for a K3 III star. Several studies of HD 81817 have been carried out since, but the origin of hybrid chromospheric features in HD 81817 is still unknown.

 We have observed HD 81817 over 15.6 years and compiled 85 high-resolution (R = 90,000) spectra as a part of our program. Since 2003 we have searched for exoplanets around K giants using the radial velocity (RV) method. The RV method is very useful in identifying companions.  Compared to main-sequence stars, however, giant stars have various stellar intrinsic mechanisms that can cause similar periodic RV variations such as rotational modulation of inhomogeneous surface features, pulsations, and stellar intrinsic activities. This makes identifying the origins of periodic RV variations in giant stars nontrivial. On the other hand, giant stars also have advantages in that they have many sharp and narrow spectral lines that facilitate precise RV measurements. Besides, giant stars have not been widely studied compare to main-sequence stars to search for exoplanets.
 
In this paper, we analyze the observed RV variations for HD 81817 and aim to identify their origins. In particular, we explore whether or not this source has a white dwarf companion, which might be an important hint to understand the nature of hybrid stars. Observations and data reduction are given in Section 2. In Section 3, we analyze the periods of RV measurements of HD 81817 to test the existence of its white dwarf companion. In Section 4, we check the photometry and chromospheric activity indices to understand the origin of RV variation periods for HD 81817. In Section 5, we analyze the IUE spectra for HD 81817. We summarize our findings in Section 6.

\section{Data}
\subsection{Observations and data reduction}

 We obtained 85 spectra for HD 81817 from May 2004 to December 2019 using the high-resolution fiber-fed Bohyunsan Observatory Echelle Spectrograph (BOES; \citealt{Kim}) on the 1.8 m telescope at Bohyunsan Optical Astronomy Observatory (BOAO), which can obtain spectrum from 3,500 {\AA} to 10,500 {\AA} with a single exposure. We used a fiber with a diameter of 80 microns that provides a resolution of R = 90,000. An iodine (I$_{2}$) cell was used for precise wavelength calibration. We used the Image Reducation and Analysis Facility (IRAF) package for raw and 1-D data reduction, RVI2CELL code (\citealt{Han2007}) for RV measurements, and Systemic Console (\citealt{Meschi}) for the analysis and fitting of RV measurements.

\subsection{Stellar model}

 We obtained fundamental photometric parameters such as spectral type, $\textit{$m_{v}$}$, and B-V for HD 81817 from the \emph{HIPPARCOS} catalog (\citealt{ESA} and \citealt{van}). We only used the parallax from the Gaia survey (Gaia DR2; \citealt{Gaia})  because the stellar atmospheric parameters from Gaia, especially $\rm{T_{eff}}$, are very different from other references. The initial parameters, especially effective temperature, play an important role in the process
of calculating the stellar parameters. We suspect that stellar atmospheric parameters for HD 81817 from Gaia DR2 are not reliable. \cite{andrae} shows the color versus temperature relations for Gaia validation data against the estimates of the effective temperatures from the literatures, and there are significant differences of about $\pm$
300 K with other references around ~4000 K. Instead of Gaia data, we derived atmospheric parameters such as $\rm{T_{eff}}$, $\mathrm{[Fe/H]}$, $v_{\mathrm{micro}}$, and log $\it g$ from the TGVIT stellar model code (\citealt{Takeda02}) using 156 equivalent widths (EWs) of Fe I and Fe II lines. We calculated stellar mass, radius, age, and log $\it g$ using the online stellar parameter estimator\footnote{http://stevoapd.inaf.it/cgi-bin/param} (\citealt{dasilva}). Rotational velocity ($v_{\mathrm{rot}}$ sin $i$) was estimated from the SPTOOL code (\citealt{Takeda08}). These methods are widely used and show a good correlation with other references. If we adopt 4043${^{+ 50}_{- 20}}$K, which is the effective temperature from Gaia DR2, then the resulting radius and mass are 93 $\pm$ 7.8 $R_{\odot}$ and 3.98 $\pm$ 0.6 $M_{\odot}$. These values are also consistent with our results within the uncertainties. The stellar parameters of HD 81817 are summarized in Table \ref{table:1}.

\begin{table}\tiny
\caption{Stellar parameters for HD 81817.} 
\label{table:1} 
\centering          
\begin{tabular}{c c c c}
\hline
Parameter &  & HD 81817 & Ref.\\\hline
Spectral type                                   &                        &K3 III           &\emph{HIPPARCOS} (\citealt{ESA})\\
$\textit{$m_{v}$}$                              & [mag]          &4.426 $\pm$ 0.001            &\cite{van}\\
$\emph{B$-$V}$                          & [mag]                  &1.488 $\pm$ 0.001            &\cite{van}\\
$\pi$                                           & [mas]                  & 3.72 $\pm$ 0.29                &\cite{Gaia}\\
$\mathrm{[Fe/H]}$                               & [dex]          &- 0.17 $\pm$ 0.1 &This work\\
                                                        &                        &       0.09            &\cite{Soubiran2016}\\
 $\rm{T_{eff}}$                         &  [K]           & 4140 $\pm$ 55                 &This work\\
log $\it g$                                     & [cgs]          &1.3 $\pm$ 0.2  &This work\\
$v_{\mathrm{micro}}$                    &[km s$^{-1}$] &2.3 $\pm$ 0.2    &This work\\
$\textit{$L_{\star}$}$          &[$L_{\odot}$] &1822.9           &This work\\
$\textit{$M_{\star}$}$                  &[$M_{\odot}$] & 4.3 $\pm$ 0.5&This work\\
$\textit{$R_{\star}$}$          &[$R_{\odot}$] &83.8 $\pm$ 7.8&This work\\
Age                                                     &[Gyr]        &0.15 $\pm$ 0.04&This work\\
 $v_{\mathrm{rot}}$ sin $i$             &[km s$^{-1}$] &4.7 $\pm$ 0.1 & This work\\
                                                        &                        &4.9            &\cite{Glebocki2005}\\
                                                        &                        &       5.5 $\pm$ 0.1&\cite{deMedeiros}\\
$P_{\mathrm{rot}}$/sin $i$              &[days]          & 801 $\sim$ 1008 & This work\\\hline
\end{tabular}
\end{table}

\section{Radial velocity measurements}
\subsection{Keplerian fit}
\begin{figure}
\resizebox{\hsize}{!}{\includegraphics{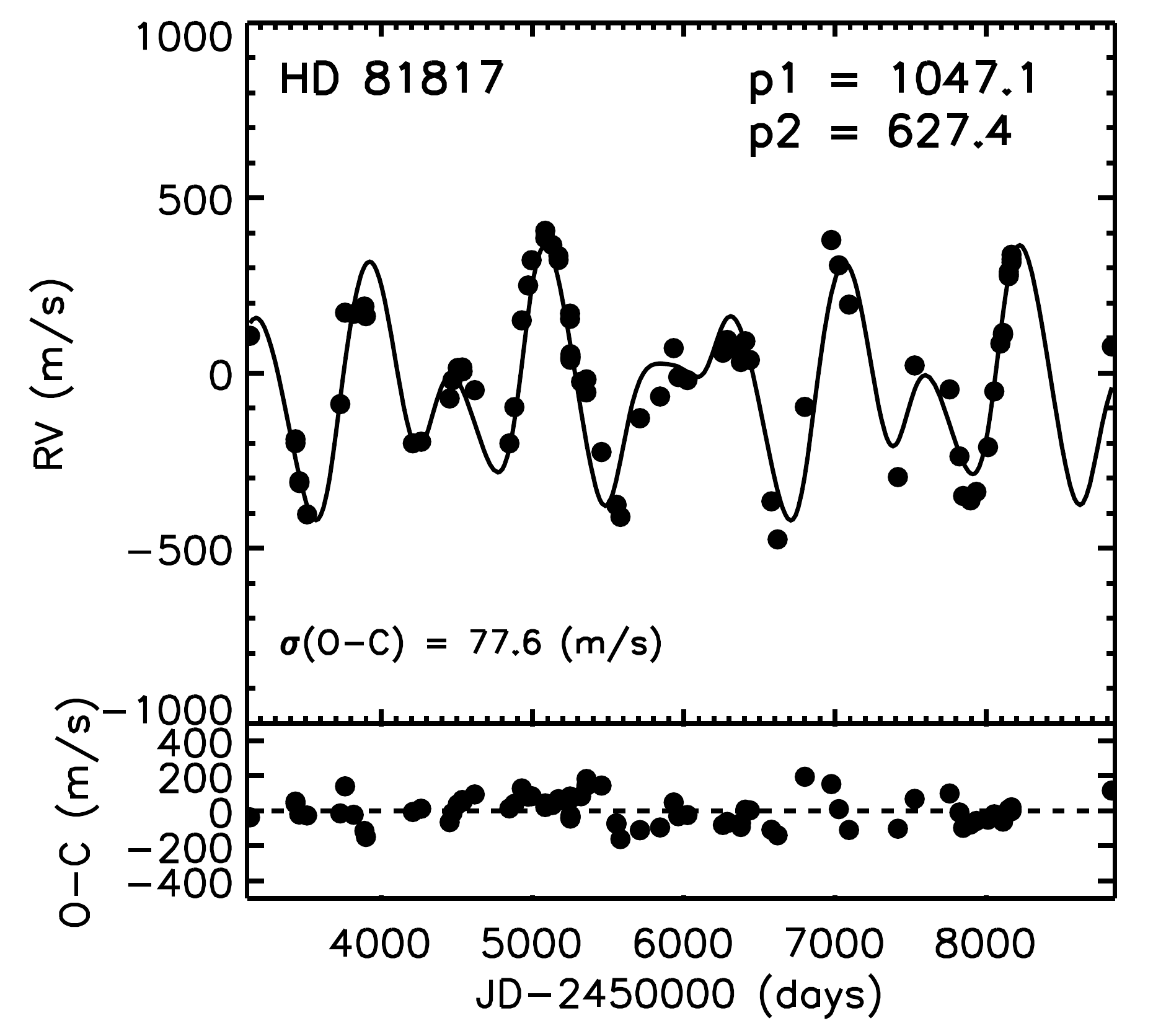}}
  \caption{RV measurements and Keplerian orbital fit (solid line) for HD 81817. There are two significant periods (1047.1 and 627.4 days) in the RV variations.}
  \label{fig1}
\end{figure}

\begin{figure}
\resizebox{\hsize}{!}{\includegraphics{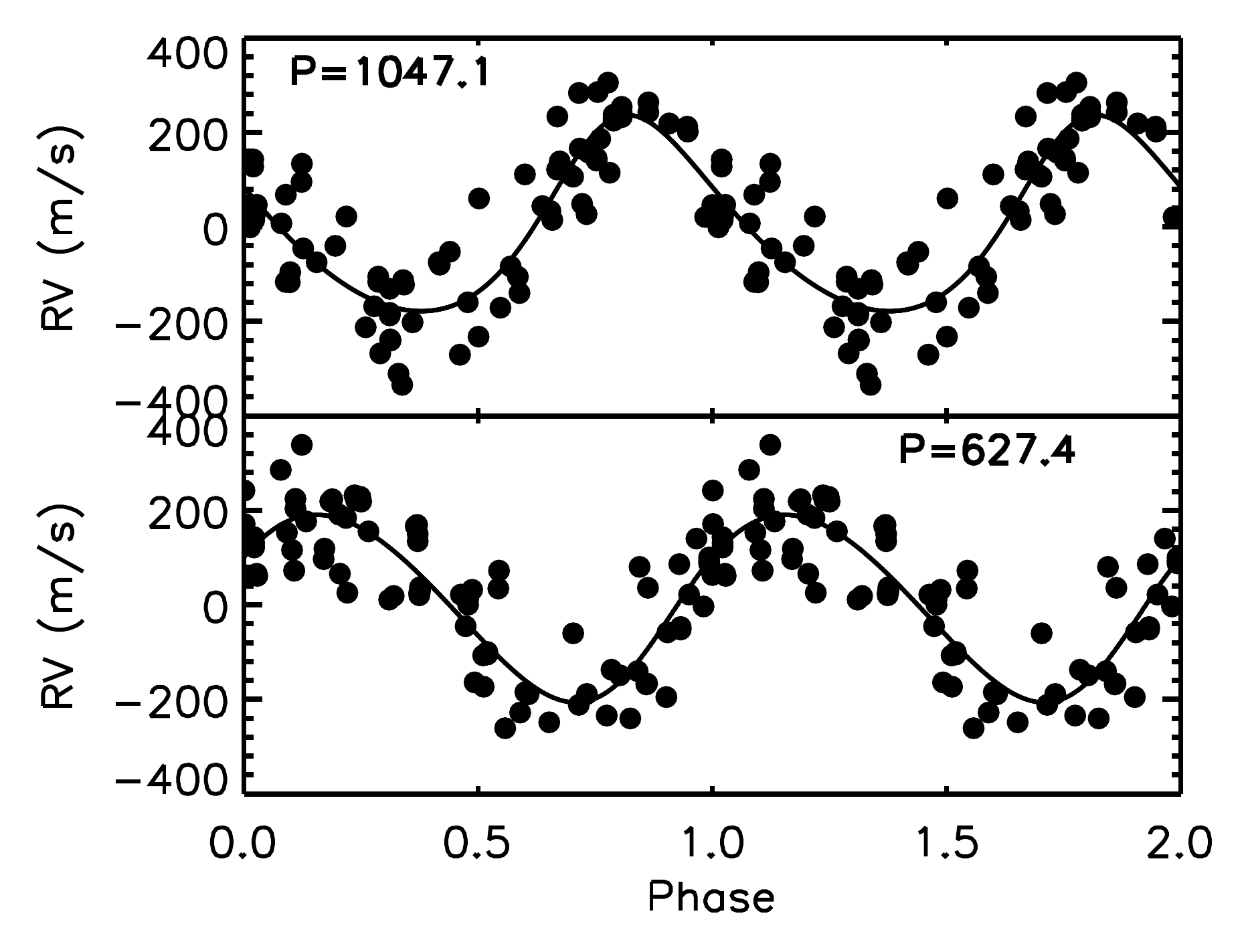}}
  \caption{Phase diagrams of two periods in RV variations for HD 81817.}
  \label{fig2}
\end{figure}

Measured RVs from the spectra of HD81817 are given in Table \ref{table:2}. We calculated the periods of RV variations using Lomb-Scargle periodograms (\citealt{Scargle}). Figure \ref{fig1} shows the RV measurements and Keplerian fitting curve (solid line) of HD 81817. Lomb-Scargle periodograms of RV show two significant periods, 1047.1 and 627.4 days with the semi-amplitudes of 207.5 m s$^{-1}$ and 198.5 m s$^{-1}$, respectively (top panel of Fig. 1). The orbital parameters from the best-fit Keplerian orbit are listed in Table \ref{table:3}. The residuals of the Keplerian fitting is 77.6 m s$^{-1}$. Giant stars, in general, show complex and irregular RV variations called jitter. The amount of RV jitter depends on the spectral type (\citealt{frink}, \citealt{hekker}, \citealt{reffert}); this value is about 50 m s$^{-1}$ for early K giants (K3 III for HD 81817). The root mean squares from our orbit fitting are compatible with these values.
Figure \ref{fig2} shows the phase diagrams for each period; these show roughly sinusoidal variations.

\begin{table}
\caption{Measurements of RVs for HD 81817 from May 2004 to December 2019.}       
\label{table:2}   
\centering                              
\begin{tabular}{crr|crr} 
\hline
JD         & $\Delta$RV  & $\pm \sigma$ &        JD & $\Delta$RV  & $\pm \sigma$  \\
$-$2,450,000 & m\,s$^{-1}$ &  m\,s$^{-1}$ & $-$2,450,000  & m\,s$^{-1}$ &  m\,s$^{-1}$  \\
\hline
3133.071        &       106.3   &       9.3     &       5554.350        &       -377.7  &       11.2    \\
3433.296        &       -200.1  &       8.1     &       5555.311        &       -376.2  &       17.1    \\
3433.362        &       -189.2  &       8.5     &       5581.151        &       -410.7  &       10.3    \\
3459.057        &       -314.1  &       8       &       5710.073        &       -129            &       18.2    \\
3459.117        &       -308.9  &       8.4     &       5710.087        &       -129            &       17.9    \\
3459.189        &       -312            &       10      &       5843.328        &       -67.1   &       8.4     \\
3510.006        &       -403.4  &       9.4     &       5933.296        &       71.1            &       11.4    \\
3729.290        &       -88.7   &       10.2    &       5963.174        &       -11.5   &       10      \\
3761.173        &       172.3   &       9       &       6023.966        &       -20.1   &       7.6     \\
3818.071        &       169             &       7.2     &       6258.380        &       57.9            &       7.8     \\
3889.016        &       189.8   &       9.8     &       6287.278        &       94.5            &       11.9    \\
3899.001        &       162.3   &       8.6     &       6288.088        &       89.2            &       11      \\
4209.019        &       -200.7  &       14.3    &       6376.969        &       31.4            &       8.5     \\
4264.055        &       -196.1  &       10.2    &       6378.007        &       52.6            &       12.6    \\
4452.377        &       -72.6   &       9.2     &       6406.989        &       90.8            &       9.9     \\
4471.278        &       -20.3   &       7.6     &       6437.002        &       37.2            &       9.5     \\
4506.180        &       14.5            &       8.2     &       6579.132        &       -366.3  &       9.2     \\
4536.088        &       15.9            &       9.6     &       6620.067        &       -475            &       10.8    \\
4538.109        &       4.9             &       9.5     &       6800.082        &       -96.6   &       11.6    \\
4618.001        &       -48.9   &       9.2     &       6975.157        &       379.7   &       8.9     \\
4847.252        &       -200.9  &       7.7     &       7024.965        &       307.5   &       11.1    \\
4880.200        &       -97.3   &       8       &       7092.140        &       195.3   &       12.2    \\
4929.051        &       150.3   &       8.4     &       7415.105        &       -297            &       8.5     \\
4971.020        &       250             &       8.8     &       7525.989        &       21.6            &       10.1    \\
4994.003        &       322.2   &       8.9     &       7757.011        &       -46.7   &       9.3     \\
5084.333        &       385.1   &       10.2    &       7821.980        &       -237.9  &       11      \\
5084.337        &       406             &       13.7    &       7846.110        &       -351            &       8.3     \\
5130.990        &       365.3   &       9.6     &       7895.045        &       -363.6  &       9.1     \\
5171.202        &       334.3   &       9.1     &       7934.003        &       -339.2  &       8.6     \\
5172.396        &       322.9   &       7.4     &       8010.943        &       -211.4  &       10.4    \\
5248.117        &       169.7   &       8.9     &       8052.341        &       -52.4   &       10.2    \\
5248.119        &       154.6   &       10.7    &       8092.312        &       84.2            &       10.1    \\
5250.196        &       43.6            &       9.7     &       8109.991        &       110.8   &       11.9    \\
5250.199        &       37.8            &       8.7     &       8109.993        &       114.6   &       12      \\
5250.201        &       41.5            &       9       &       8109.995        &       109.3   &       12.7    \\
5251.089        &       47.4            &       10.2    &       8148.078        &       276.5   &       10.2    \\
5251.091        &       53.8            &       9.4     &       8148.085        &       289.1   &       10.5    \\
5251.092        &       45.4            &       8.8     &       8148.093        &       281.6   &       9.3     \\
5321.000        &       -24.9   &       8.9     &       8166.128        &       337.6   &       10.3    \\
5356.090        &       -55.1   &       10.7    &       8166.136        &       316             &       10.5    \\
5357.006        &       -18.1   &       9.3     &       8166.143        &       324.9   &       11      \\
5456.343        &       -225.5  &       9.7     &       8829.206        &       75.9            &       10.5    \\
5554.343        &       -376.1  &       11.4    \\                                              
\hline
\end{tabular}
\end{table}

\begin{table}\large
\caption{Orbital parameters for HD 81817 b} 
\label{table:3} 
\centering          
\begin{tabular}{c|c|c}  
\hline
Parameter& &HD 81817 b\\\hline
P &[days]& 1047.1 $\pm$ 8.5\\
$\it T$$_{\rm{periastron}}$&[JD] &2449712 $\pm$ 108 \\
$\it{K}$& [m s$^{-1}$]&211.4 $\pm$ 16.4\\
$\it{e}$ && 0.17 $\pm$ 0.07 \\
$\omega$ &[deg]& 320 $\pm$ 34\\
 rms &[m s$^{-1}$]&77.6\\\hline
\emph{m}~sin~$\it i$& [$\rm{M_{Jup}}$] &27.1 $\pm$ 2.1\\
$\it{a}$ &[AU] &3.3 $\pm$ 0.1\\\hline
\end{tabular}
\end{table}

\subsection{Testing existence of white dwarf companion}

Despite the suggestion that HD 81817 has a white dwarf companion, there have been no kinematic studies to prove or disprove the existence of a white dwarf companion. Our RV measurements of HD 81817 show periodic variations and the amplitude for two RV variation periods were around 200 m s$^{-1}$. To test the existence of a white dwarf companion, we calculated the RV variation expected from a white dwarf companion with the typical mass of 0.7 $M_{\odot}$, which has an orbital period of 15.6 years (the duration of our observation). The calculated amplitude is about 1.3 $\sin i$ km s$^{-1}$; this value is significantly higher than our observed RV amplitudes, unless $\sin i$ is very small. The inclination of this system should be about 8.8$^\circ$ or less. The probability for such a small inclination angle is less than 1.2\%. In addition, the unprojected rotational velocity of HD 81817 is about 31 km s$^{-1}$.
Considering the spectral type and radius of HD 81817, this value is unrealistic because these stars with this spectral type have much lower rotational velocities, that is, lower than 10 km/s (\citealt{fekel}).

The secular linear change in RV over 15.6 years is about 7.96 km s$^{-1}$ yr$^{-1}$. Any hypothetical companion with mass 0.7 $M_{\odot}$ should have an orbital period that is much longer than 15.6 years, unless the orbit is very eccentric and the viewing geometry is special. The expected maximum angular separation between HD 81817 and a 0.7 $M_{\odot}$ white dwarf companion with an orbital period of 15.6 years is at least larger than 0.04 arcsec, which is almost the same as the angular resolution of the Hubble Space Telescope (\emph{HST)}. In the absence of a \emph{HST} observation on HD 81817, we cannot completely rule out a very distant white dwarf companion with our RV data alone.

Recently, \cite{kervella} has shown that HD 81817 has a companion with a mass of about 124${^{+ 97.39}_{- 91.97}}$ $\rm{M_{Jup}}$ and semimajor axis of 2.67 AU based on Gaia DR2 proper motion anomaly. Considering the large uncertainties in mass estimate, the result is consistent with our Keplerian fit. Therefore, the existence of a substellar companion at $\sim$ 3 AU for HD 81817 seems secure.

\section{Origins of RV variations}

Determining the origin of observed periodic RV variations in giant stars is nontrivial, but essential, for stellar and exoplanetary studies. We analyzed the \emph{HIPPARCOS} photometric data and stellar activity indices such as Ca II H lines, bisectors, $\rm{H_{\alpha}}$, and $\rm{H_{\beta}}$ EWs.

\begin{figure}
\resizebox{\hsize}{!}{\includegraphics{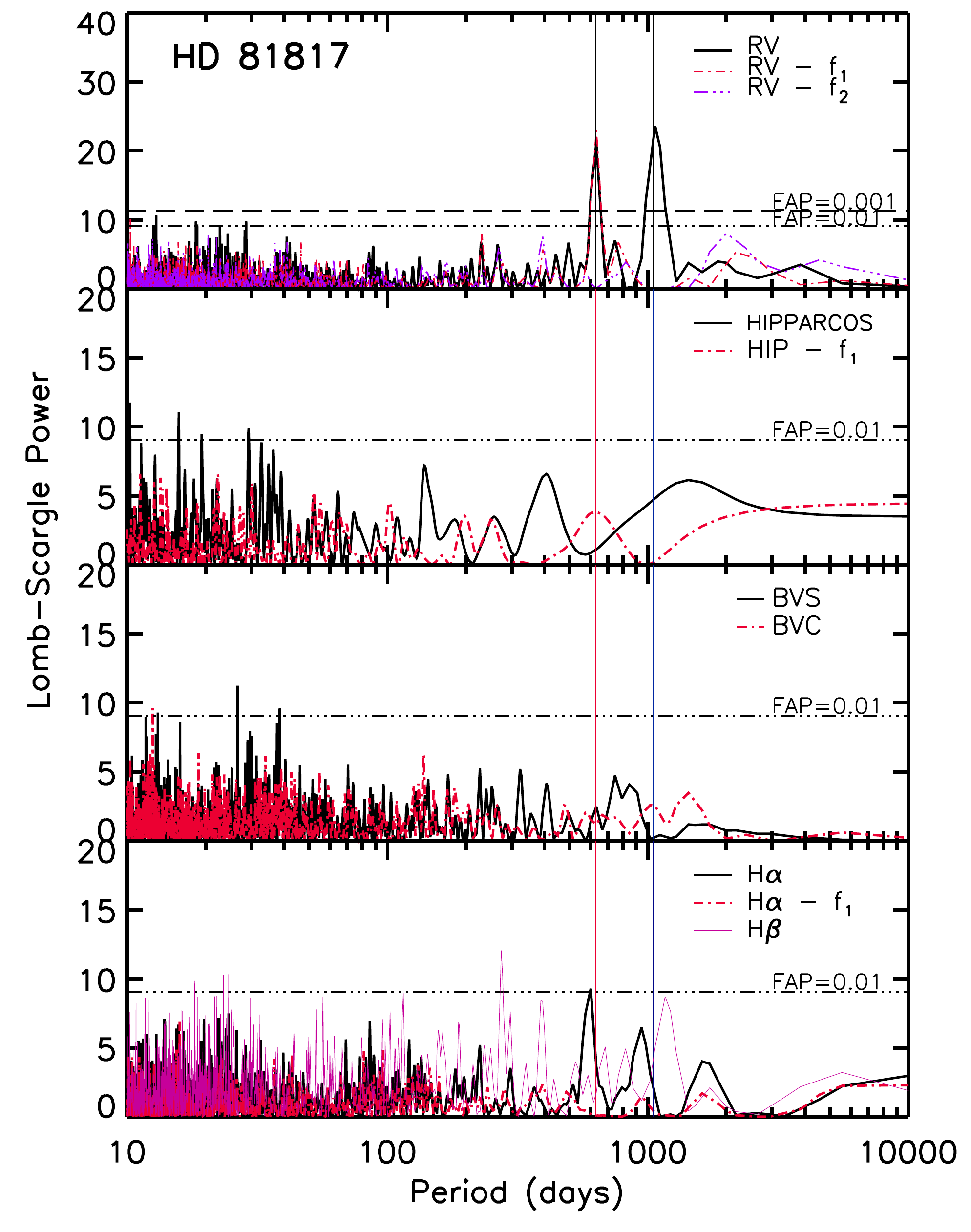}}
  \caption{Lomb-Scargle periodogram of RV measurement, \emph{HIPPARCOS}, bisectors, $\rm{H_{\alpha}}$, and $\rm{H_{\beta}}$ EW measurements (from top to bottom) for HD 81817. Vertical lines show two periods of 1047.1 days and 627.4 days in the RV variations larger than the power of FAP of 0.01.}
  \label{fig3}
\end{figure}

\subsection{HIPPARCOS photometry}

We checked the variation in the photometry data from \emph{HIPPARCOS}. Stellar pulsations can produce photometric variations as well as RV variations. If pulsations cause both RV and photometric variations, then the periods of the photometric and RV variations overlap in the Lomb-Scargle periodogram. Although HIPPARCOS data were not contemporaneous with our RV measurements, the same periodicities appear both in the photometry and RV variations of HD 81817. The Lomb-Scargle periodogram does not show any significant periods in the photometric variations that coincide with RV variation periods as shown in Figure \ref{fig3} (first and second panels). Thus, RV variations with periods of 1047.1 days and 627.4 days are not accompanied by photometric variations.

\subsection{Chromospheric activity}

\begin{figure}
\resizebox{\hsize}{!}{\includegraphics{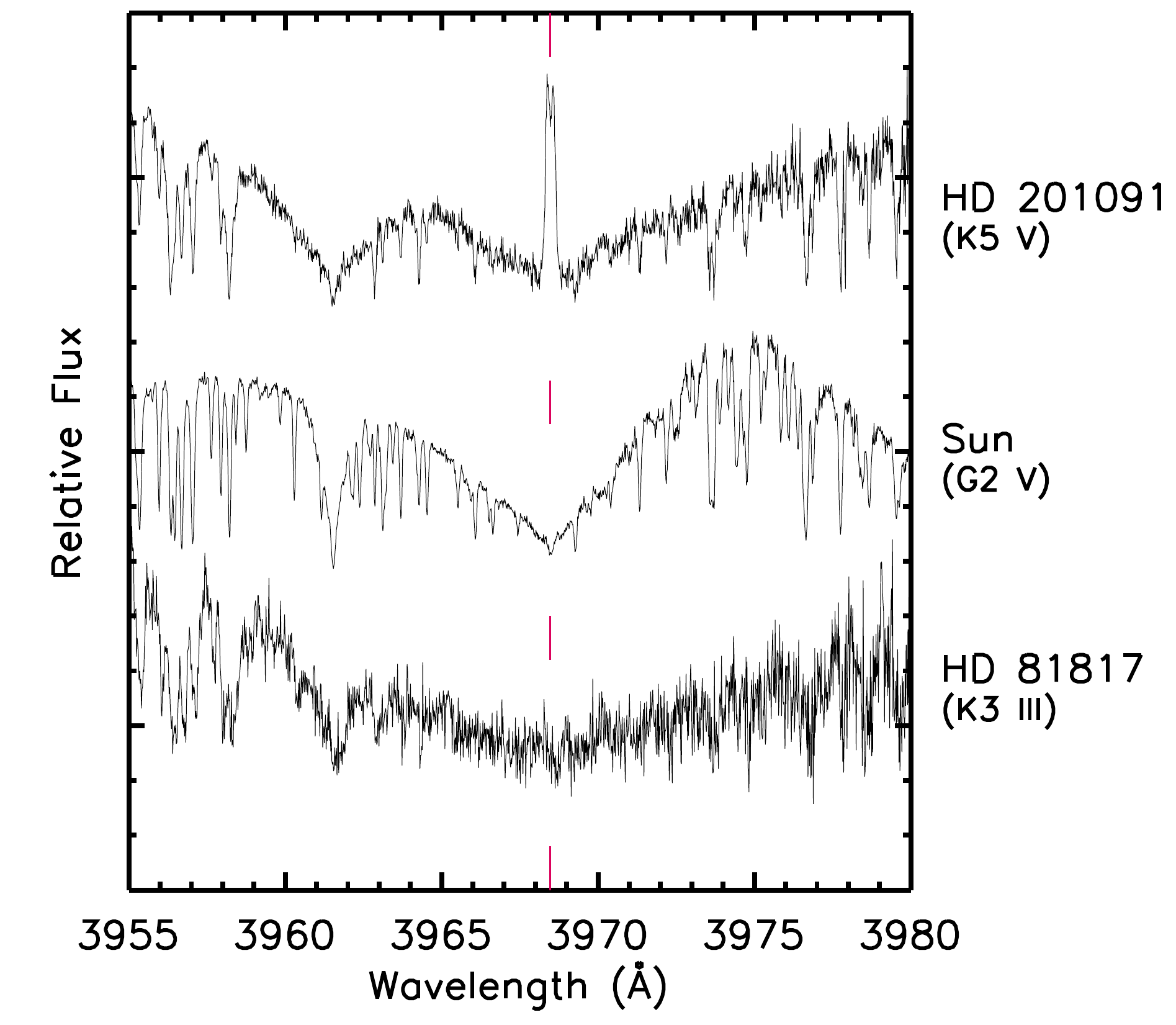}}
  \caption{Spectra of Ca II H lines for chromospheric active star HD 201091, Sun, and HD 81817 (from top to bottom). Dashed line shows the center of Ca II H line.}
  \label{fig4}
\end{figure}

The Ca II H line is a representative indicator of stellar chromospheric activity discovered by \cite{Eberhard}. Chromospheric activities in the star can make strong emission features at the center of the Ca II H line. Figure \ref{fig4} shows Ca II H lines of the chromospherically active star HD 201091, the Sun, and HD 81817. Our Ca II line profile of HD 81817 is rather noisy, but does not show any prominent core emission features.

We examined other stellar activity indicators, $\rm{H_{\alpha}}$ and $\rm{H_{\beta}}$ EW variations. The bottom panel of Figure \ref{fig3} shows the Lomb-Scargle periodogram of $\rm{H_{\alpha}}$ and $\rm{H_{\beta}}$ EW variations for HD 81817. In the periodogram of $\rm{H_{\alpha}}$ EW variations, we find a significant period of  $\sim$600 days with a false alarm probability (FAP) of 0.01, which is similar to the RV period of 627.4 days. We also find a significant period of $\sim$273 days for $\rm{H_{\beta}}$ EW variations. However, we do not find any periodicities in the Lomb-Scargle periodograms of other parameters, which seem to be related to the period of $\sim$ 273 days for $\rm{H_{\beta}}$ EW variations. The period of $\sim$1100 days for $\rm{H_{\beta}}$ EW variations is the harmonic of the primary period of $\sim$273 days because it disappears when the primary period is removed. Thus we conclude that $\rm{H_{\beta}}$ EW variations are not related to the observed RV variations. Based on these activity indicators, we suspect that stellar activities with a period of $\sim$600 days are likely to be related to the 627.4 days of RV variations.

\subsection{Rotational modulation of surface features}
\subsubsection{Rotational period}

Rotational periods of typical giant stars span several hundred days. The rotation of stars with an inhomogeneous surface can produce periodic RV variations. Thus, a comparison of the rotational period against the RV variation period can help determine the nature of observed RV variations. We calculated the rotational velocity $v_{\mathrm{rot}}$ sin $i$ for HD 81817 using the SPTOOL code (\citealt{Takeda08}), which calculates the line broadening by stellar rotation using five elements (V, Ti, Fe, Ni, and Co) in the wavelength range from 6080 {\AA} to 6090 {\AA}. The calculated projected rotational velocity of HD 81817 is 4.7 km s$^{-1}$, which is very consistent with other references (see Table \ref{table:1}). We estimate that the rotational period of HD 81817 ranges from 801 $\sin i$ to 1008 $\sin i$ days (68\% CL). If we adopt Gaia stellar parameters, however, the rotational period ranges from 898 $\sin i$ days to 1085 $\sin i$ days (68\% CL). The probability for $\sin i$ greater than 1047/1085 is less than 7\%. Hence, RV variations of period 1047 days are unlikely to be related to the rotational modulation, albeit this is not impossible.

\subsubsection{Bisectors}

\begin{figure}
\resizebox{\hsize}{!}{\includegraphics{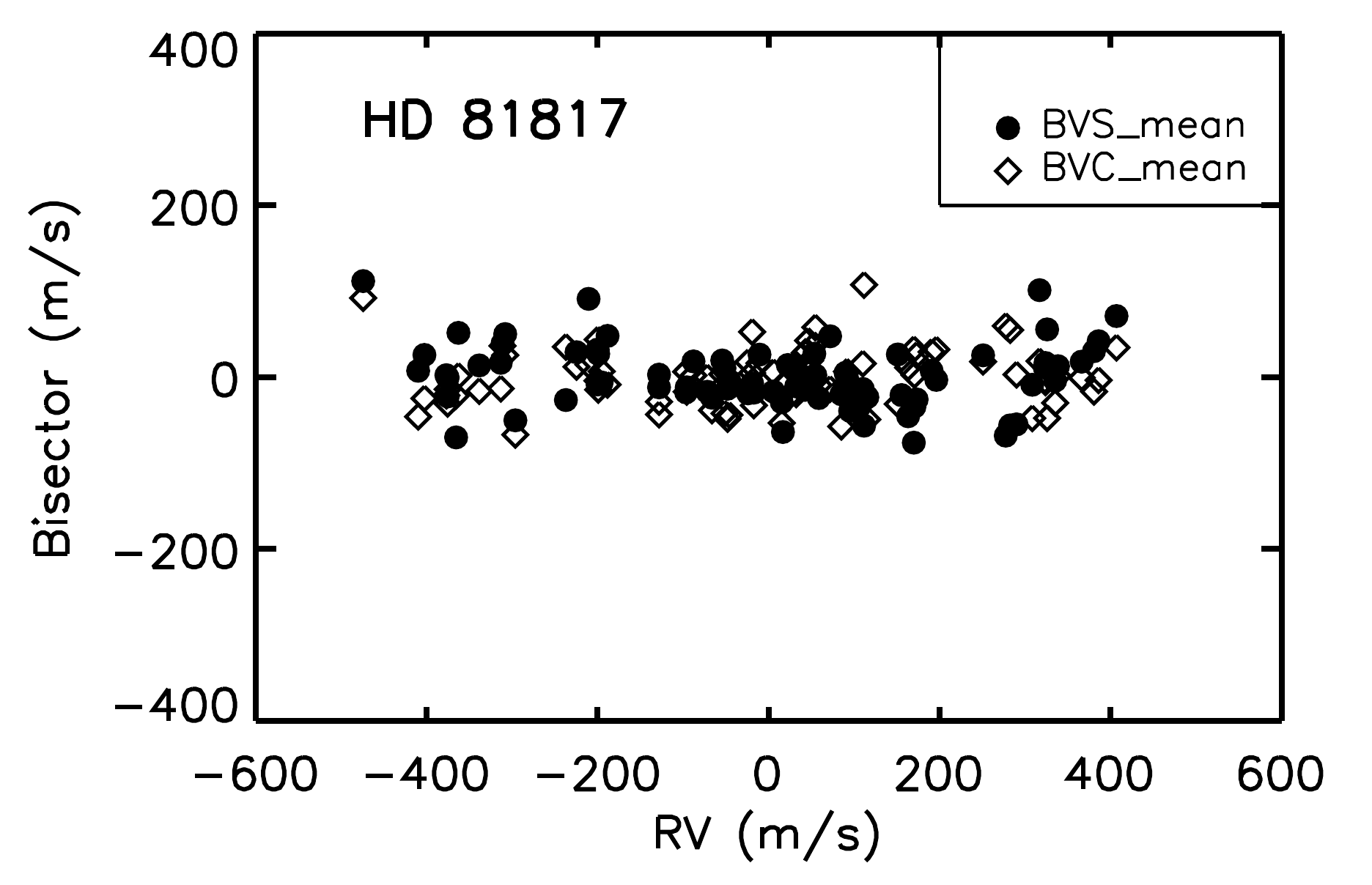}}
  \caption{Correlations between RV measurements and mean values of BVS (filled circles) and BVC (open diamonds). There are no significant correlations.}
  \label{fig5}
\end{figure}

Rotational modulation of surface features like starspots can produce spectral line asymmetries, which can be quantified by bisectors (\citealt{Hatzes1996}; \citealt{Hatzes1998}). The bisector velocity span (BVS) is defined as the difference in line centers between the high and low levels of the line profile. Bisector velocity curvature (BVC) is the difference in the velocity span of the upper half and  lower half of the line profile. We averaged bisectors of five sharp and narrow absorption lines for HD 81817 (Sc I 6210.6 {\AA}, Fe I 6261.1 {\AA}, Fe I 6546.2 {\AA}, Ca I 6572.8 {\AA}, and Ni I 6767.8 {\AA}). We checked the Lomb-Scargle periodogram of BVSs and BVCs (third panel of Figure \ref{fig3}). No significant periods exist that match those of RV variations. We also checked the correlations between two bisector parameters and RV measurements. Figure \ref{fig5} shows BVSs  and BVCs  against RV values for HD 81817. There are no correlations between bisectors and RV measurements. Thus we conclude that the rotational modulation of surface features is not the cause of periodic RV variations.

\section{Reanalysis of the IUE spectra}
\begin{figure}
\resizebox{\hsize}{!}{\includegraphics{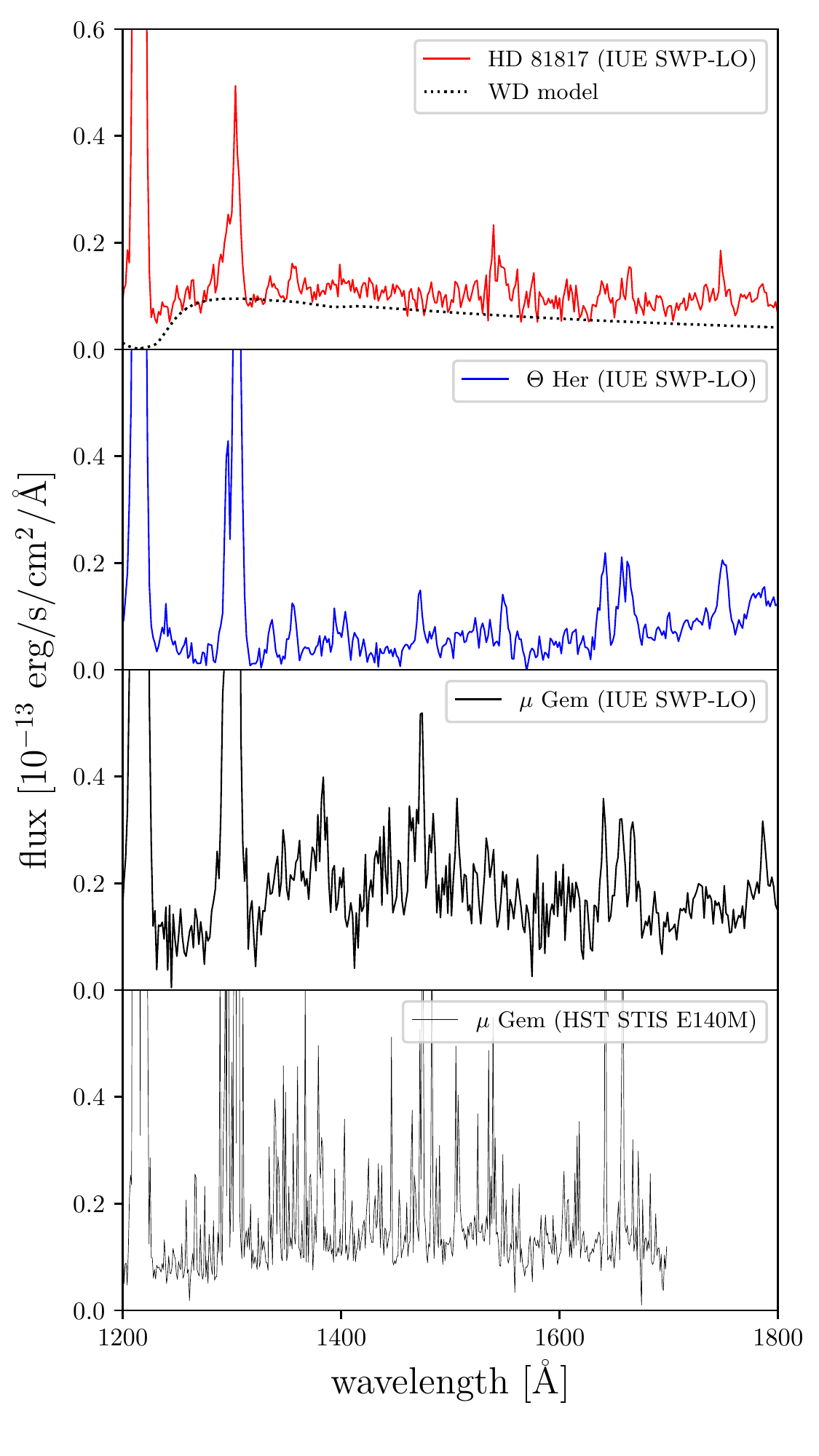}}
  \caption{Ultraviolet spectra of HD 81817, $\Theta$~Her, and $\mu$~Gem. {\it Top panel}: Co-added IUE spectra of HD 81817 (observation IDs SWP19602+SWP21111, red line) and a white dwarf model from \cite{Koester} with the parameters given by \cite{Reimers84}. {\it Second panel:} Co-added IUE spectra of $\Theta$~Her (observation IDs SWP13642+SWP19624). {\it Third panel:} IUE spectrum of $\mu$~Gem. {\it Bottom panel:} HST STIS/FUV-MAMA E140M spectrum of $\mu$~Gem. The three spectra (HST observing IDs oc7x09020, oc7x09030 and oc7x09040) were co-added and rebinned with a bin size of $1$~\AA.}
  \label{fig6}
\end{figure}
Since our analysis of the RV data and GAIA data indicate that the companion to HD 81817 proposed by \cite{Reimers84} is unlikely to be a closely orbiting white dwarf, the excess flux in the IUE spectrum can either stem from an object in the foreground, background, or in distant orbit, or it can originate in the extended chromosphere of HD 81817. Unfortunately there are no UV spectra of HD 81817 acquired with the HST that would allow for a definite conclusion of this matter. We do, however, want to discuss the latter possibility by providing an example of an evolved star with a significant UV continuum.\\
For this purpose we downloaded and co-added spectra of HD~81817 and $\Theta$~Her from the INES\footnote{\url{http://sdc.cab.inta-csic.es/ines/index2.html}} archive \citep{Rodriguez} and processed these data in the same manner as described by \cite{Reimers84} (see first and second panel of Figure \ref{fig6}). In comparison to $\Theta$~Her, the spectrum of HD~81817 exhibits a clear continuum, which is described well by a white dwarf model \citep{Koester} with an effective temperature of $\rm{T_{eff}}=20\,000$~K and a surface gravity of log~$g=8$, as proposed by \cite{Reimers84}.\\
The bottom two panels of Figure \ref{fig6} show the UV spectra of $\mu$~Gem, which is a giant of spectral type M3IIIab that has been observed both by IUE and HST STIS/FUV-MAMA \citep[e.g.,][]{riley}. The low-resolution IUE spectrum (third panel) also exhibits a continuum that seems to decline toward longer wavelengths, albeit not in as smooth a fashion as that of HD~81817. The high-resolution spectrum obtained with the HST (botton panel) reveals that this effect is caused by a combination of a constant flux level of $\approx10^{-14}$~erg/s/cm$^2$/\AA~with a large number of emission lines when degraded to the lower resolution of IUE.\\
We thus conclude that to differentiate between the white dwarf hypothesis and the chromosphere or wind of HD~81817 as the true origin of the UV continuum, high-resolution UV spectra of HD~81817 are required.

\section{Summary}
\begin{figure}
\resizebox{\hsize}{!}{\includegraphics{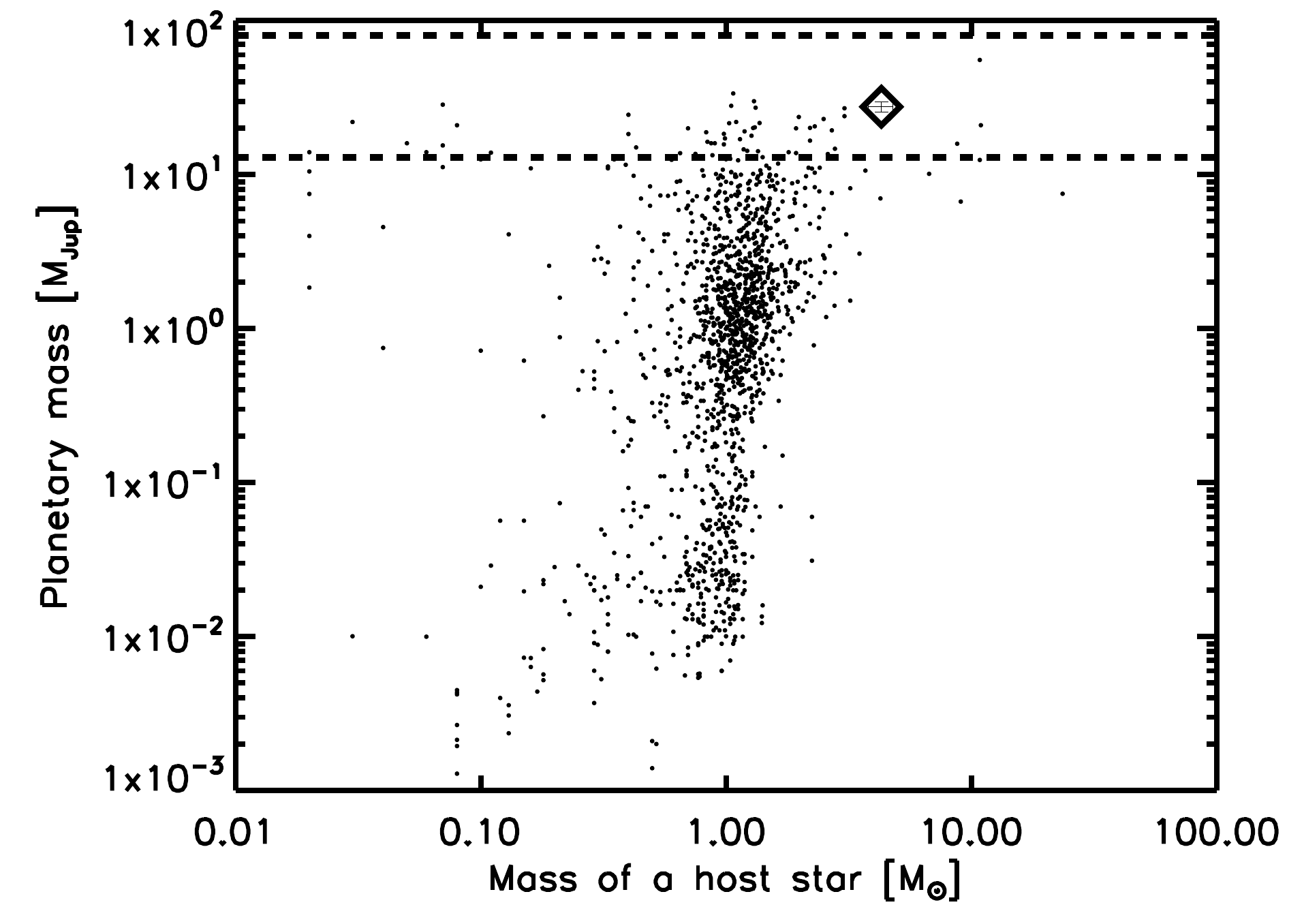}}
  \caption{Distribution of mass of planetary companions vs. stellar mass as of February 2019. The area between the dashed lines shows brown dwarf mass range, from 13 $\rm{M_{Jup}}$ to 80 $\rm{M_{Jup}}$. The black points are known companions and the diamond is  a brown dwarf or substellar companion candidate of HD 81817 b in this work.}
  \label{fig7}
\end{figure}

Hybrid stars are rare objects whose spectra have both cool stellar wind properties and UV or even X-ray emission features of highly ionized atoms. HD 81817 was identified as a hybrid star and believed to have a white dwarf companion based on its UV spectra. However, the existence of a white dwarf companion has not been tested with RV observations.

We found two significant periods in RV variations for HD 81817: the FAP of the 1047.1-day period is about 9 x 10$^{-11}$ and that of 627.4-day period about 2.5 x 10$^{-9}$. The observed amplitudes of RV variations for HD 81817 are too small compared to typical amplitudes in RV variations expected from a closely orbiting white dwarf companion.

The Lomb-Scargle periodogram does not show significant periodic photometric variations. We do not find any core emission features at the center of Ca II H line for HD 81817. Analysis of atmospheric activity indicator lines do not show obvious atmospheric activities. However, there is an  approximately 600-day period in $\rm{H_{\alpha}}$ EW variations, which may be related to the period of 627 days. Thus the 627-day RV variations are likely to be the result of rotational modulation of surface features or of long-period pulsations, while 1047-day RV variations are from an orbiting companion. Recent Gaia proper motion anomaly, which is understood to be a substellar companion at 2.67 AU (\citealt{kervella}), corroborates the latter conclusion.

Thus we conclude that 1047.1-day RV variations are caused by an orbiting brown dwarf or substellar companion, which has a minimum mass of 27.1 $M_\mathrm{Jup}$, a semimajor axis of 3.3 AU, and an eccentricity of 0.17, assuming the stellar mass of 4.3 $M_{\odot}$ for HD 81817. Especially, the inferred mass puts HD 81817 b in the brown dwarf desert (\citealt{grether}), spanning from 13 $M_\mathrm{Jup}$ to 80 $M_\mathrm{Jup}$. Figure \ref{fig7} shows the distribution of confirmed exoplanets from NASA exoplanets archive so far in the diagram for the mass of host stars versus the mass of planetary companions. HD 81817 is one of the most massive and largest stars so far to host a brown dwarf candidate or subsstellar companion and a hybrid star at the same time.

\begin{acknowledgements}

We thank the anonymous referee for useful comments. TYB and MGP were supported by the Basic Science Research Program through the National Research Foundation of Korea (NRF) funded by the Ministry of Education (2019R1I1A3A02062242), BK21 Plus of National Research Foundation of Korea (22A20130000179), and KASI under the R\&D program supervised by the Ministry of Science, ICT and Future Planning and by the NRF to the Center for Galaxy Evolution Research (No. 2017R1A5A1070354). BCL acknowledges partial support by Korea Astronomy and Space Science Institute (KASI) grant  2019-18-30-03. VP acknowledges funding by the German Aerospace Center (DLR). Parts of this publication are based on INES data from the IUE satellite. Based on observations made with the NASA/ESA Hubble Space Telescope, obtained from the Data Archive at the Space Telescope Science Institute, which is operated by the Association of Universities for Research in Astronomy, Inc., under NASA contract NAS 5-26555. Support for StarCAT was provided by grant HST-AR-10638.01-A from STScI, and grant NAG5-13058 from NASA. The project has made use of public databases hosted by SIMBAD and VizieR, both maintained by CDS, Strasbourg, France.

\end{acknowledgements}

%

\begin{thebibliography}{}


 \bibitem[Andrae et al. (2018)]{andrae} Andrae, R., Fouesneau, M., Creevey, O., et al. \ 2018, A\&A, 616, 8

     \bibitem[Ayres (2005)]{ayres1} Ayres, T. R. \ 2005, ApJ, 618, 493

     \bibitem[Ayres et al. (2006)]{ayres3} Ayres, T. R., Brown, A., \& Haper, G. M. \ 2006, ApJ, 651, 1126
      	
     \bibitem[Ayres (2011)]{ayres2} Ayres, T. R. \ 2011, ApJ, 738, 120
     	
    \bibitem[Bil{\'i}kov{\'a} et al. (2010)]{bilikova} Bil{\'i}kov{\'a}, J., Chu, You-Hua, Gruendl, R. A., \& Maddox, L. A. \ 2010, AJ, 140, 1433

    \bibitem[Caillault (1996)]{caillault} Caillault, J.-P. \ 1996, ASPC, 109, 325

    \bibitem[da Silva (2006)]{dasilva} da Silva, L., Girardi, L., Pasquini, L., et al. \ 2006, A\&A, 458, 609

   \bibitem[de Medeiros \& Mayor (1999)]{deMedeiros}de Medeiros, J. R., \& Mayor, M. \ 1999, A\&AS, 139, 433

     \bibitem[Eberhard \& Schwarzschild (1913)]{Eberhard} Eberhard, G., \& Schwarzschild, K. \ 1913, ApJ, 38, 292

   \bibitem[ESA (1997)]{ESA} ESA 1997, VizieR Online Data Catalog, 1239, 0

   \bibitem[Fekel (1997)]{fekel} Fekel, F. C. \ 1997, PASP, 109, 514

   \bibitem[Frink et al. (2001)]{frink} Frink, S., Quirrenbach, A., Fischer, D., R{\"o}ser, S., \& Schilbach, E. \ 2001, PASP, 113, 173

   \bibitem[Gaia collaboration (2018)]{Gaia} Gaia collaboration (Brown, A. G. A., et al.) \ 2018, A\&A, 616, 1

   \bibitem[Gł\c{e}bocki \& Gnaci{\'n}ski (2005)]{Glebocki2005} Gł\c{e}bocki, R., \& Gnaci{\'n}ski, P. \ 2005, ESASP, 560, 571 

   \bibitem[Grether \& Lineweaver (2006)]{grether} Grether, D., \& Lineweaver, C., H. \ 2006, ApJ, 640, 1051

   \bibitem[Haisch et al. (1992)]{Haisch} Haisch, B., Schmitt, J. H. M. M., \& Rosso, C. \ 1992, ApJ, 388, 61

   \bibitem[Han et al. (2007)]{Han2007} Han, I., Kim, K.-M., Lee, B.-C., \& Valyavin, G.\ 2007, PKAS, 22, 75

   \bibitem[Hartmann et al. (1980)]{Hartmann1980} Hartmann, L., Dupree, A.K., \& Raymond, J. C. \ 1980, ApJ, 236, 143

   \bibitem[Hartmann et al. (1981)]{Hartmann1981} Hartmann, L., Dupree, A.K., \& Raymond, J. C. \ 1981, ApJ, 246, 193

   \bibitem[Hatzes (1996)]{Hatzes1996} Hatzes, A.~P. \ 1996, PASP, 108, 839

   \bibitem[Hatzes et al. (1998)]{Hatzes1998} Hatzes, A.~P., Cochran, W.~D., \& Bakker, E.~J.\ 1998, Nature, 391, 154

   \bibitem[Hekker et al. (2006)]{hekker} Hekker, S., Reffert, S., Quirrenbach, A., et al. \ 2006, A\&A, 454, 943

   \bibitem[Jeong et al. (2018)]{Jeong2018} Jeong, G., Lee, B.-C., Han, I., et al. \ 2018, A\&A, 610, 3

   \bibitem[Kervella et al. (2019)]{kervella} Kervella, P., Arenou, F., Mignard, F., \& Th{\`e}venin, F. \ 2019, A\&A, 623, 72

   \bibitem[Kim et al. (2007)]{Kim} Kim, K.-M., Han, I., Valyavin, G.~G., et al.\ 2007, PASP, 119, 1052

   \bibitem[Koester (2010)]{Koester} Koester, D. \ 2010, MmSAI, 81, 921

   \bibitem[Meschiari et al. (2009)]{Meschi} Meschiari, S., Wolf, A. S., Rivera, E., et al. \ 2009, PASP, 121, 1016

   \bibitem[Reffert et al. (2018)]{reffert} Reffert, S., Bergmann, C., Quirrenbach, A., Trifonov T., \& K{\"u}nstler, A. \ 2015, A\&A, 574, 116

   \bibitem[Reimers (1982)]{Reimers82} Reimers, D. \ 1982, A\&A, 107, 292

   \bibitem[Reimers (1984)]{Reimers84} Reimers, D. \ 1984, A\&A, 136, 5

   \bibitem[Reimers \& Schmitt (1992)]{Reimers1992} Reimers D., \& Schmitt, J. H. M. M. \ 1992, ApJ, 392, 55

   \bibitem[Riley (2017)]{riley} Riley, A. \ 2017, STIS Instrument Hanbook for Cycle 25, Version 16.0

   \bibitem[Rodr{\'{\i}}guez-Pascual et al. (1999)]{Rodriguez} Rodr{\'{\i}}guez-Pascual, P. M., Gonz{\'a}lez-Riestra, R., Schartel, N., \& Wamsteker, W. \ 1999, A\&AS, 139, 183

   \bibitem[Scargle (1982)]{Scargle} Scargle, J.~D.\ 1982, ApJ, 263, 835

   \bibitem[Soubiran et al. (2016)]{Soubiran2016} Soubiran, C., Le Campion, J.-F., Brouillet, N., \& Chemin, L., \ 2016, A\&A, 591, 118

   \bibitem[Takeda et al. (2002)]{Takeda02} Takeda, Y., Ohkubo, M., \& Sadakane, K. \ 2002, PASJ, 54, 451

   \bibitem[Takeda et al. (2008)]{Takeda08} Takeda, Y., Sato, B., \& Murata, D. \ 2008, PASJ, 60, 781

   \bibitem[van Leeuwen (2007)]{van} van Leeuwen, F. \ 2007, A\&A, 474, 653

































\end{thebibliography}
%
\bibliographystyle{plain}

\end{document}